\documentclass[reprint,twocolumn,superscriptaddress,amsmath,amssymb,aps,pre]{revtex4-1}

\usepackage{graphicx}
\usepackage{graphics}
\usepackage{amsmath}
\usepackage{dcolumn}
\usepackage{bm}

\begin{document}


\title{Field driven phase transition from semiconductor to half-metallic ferrimagnet of TcO$_2$ uni-cell layer on rutile TiO$_2$(001) surface}

\author{Xiang-Bo Xiao}
\affiliation{Beijing National Laboratory for Condensed Matter Physics, Institute of Physics, Chinese Academy of Sciences, Beijing 100190, China}
\affiliation{School of Physical Sciences, University of Chinese Academy of Sciences, Beijing 100190, China}
\author{Bang-Gui Liu}\email{bgliu@iphy.ac.cn}
\affiliation{Beijing National Laboratory for Condensed Matter Physics, Institute of Physics, Chinese Academy of Sciences, Beijing 100190, China}
\affiliation{School of Physical Sciences, University of Chinese Academy of Sciences, Beijing 100190, China}

\date{\today}

\begin{abstract}
For spintronics applications, it is highly desirable to realize highly-spin-polarized two-dimensional (2D) electron systems in electrically-controllable epitaxial ultrathin films on semiconductor substrates. Through systematic first-principles investigation, we propose the TcO$_2$ uni-cell layer (one-unit-cell thickness) on rutile TiO$_2$ (001) substrate as a semiconductor heterostructure and use electric field to manipulate its electronic and magnetic properties. Our study shows that the heterostructure is a narrow-gap semiconductor with an antiferromagnet-like ordering when the applied electric field is less than 0.026 V/\AA{}, and then it transits to a half-metallic ferrimagnet with 100\% spin polarization. Our further analysis indicates that the magnetization density and the electronic states near the Fermi level originate mainly from the TcO$_2$ uni-cell layer, with the remaining minor part from the interfacial Ti-O$_2$ monolayers, and the bonds and bond angles quickly converge to the corresponding values of bulk TiO$_2$ when crossing the interface and entering the TiO$_2$ layer. Therefore, the heterostructure is actually a 2D electron system determined by the TcO$_2$ uni-cell layer and the TiO$_2$ substrate. Because the half-metallic phase with 100\% spin polarization can be achieved at 0.026 V/\AA{}, this epitaxial 2D electron system should be usable in spintronics applications.
\end{abstract}

\pacs{Valid PACS appear here}
\maketitle


\section{Introduction}

Semiconductor heterostructures have attracted much attention because controllable interfaces and excellent two-dimensional (2D) electron systems can be obtained \cite{sh1,sh2,sh3,sh4,sh5,sh6,sh7,addq,addq2,addq4}. It is always exciting to seek new spintronic materials for functional devices from semiconductor heterostructures, because the spintronic applications need combining the full (100\%) spin polarization of carriers with modern semiconductor technology \cite{ss1,ss2,ss3,addq1,addq3}. Half-metal is ideal spintronic material because of its full spin polarization, thanks to its key feature that one of the spin channel is semiconductor and the other is metallic \cite{hm1,adlbg1,adlbg2,adlbg3}. For example, rutile CrO$_2$, a distinguished half-metal, has been intensively investigated theoretically and experimentally \cite{co1,co2,co3,co4,co5,co6,co7,adxxb}.
TiO$_2$ is a semiconductor with a gap around 3.0 eV. It could be used to achieve  potential photocatalyst materials, solar energy materials, dilute magnetic semiconductors, and so on \cite{tio1,tio2,tio3,tio4,tio5,tio6,tio7,tio8,tio9}. Interestingly, some effects of electric field-induced resistive switching  have been observed in oxide TiO$_2$ \cite{rs1,rs2,rs3,rs4}. Recently, Technetium (Tc) based antiferromagnetic perovskites CaTcO$_3$ and SrTcO$_3$ became attractive because of thir high Neel temperatures, exceeding 800 K and 1000 K, respectively \cite{stco1,ctco1}. The high Neel temperatures can be understood by combining the cooperative rotation of the TcO$_6$ octahedrons and the itinerant-to-localized transitions in the Tc-based compounds \cite{stc1,stc2}. TcO$_2$ is also very interesting because of its similarity to SrTcO$_3$. It is highly desirable to realize highly-spin-polarized 2D electron systems in epitaxial ultrathin Tc-based films on appropriate semiconductor substrates. At the same time, the electrically-controlled magnetism or electric-field-driven magnetic phase transitions, especially accompanying some semiconductor-metal transitions, can play an important role in potential applications in spintronics \cite{ecm1,ecm2,ecm3}.

Here, we design a semiconductor heterostructure consisting of one TcO$_2$ uni-cell  epitaxial layer on the rutile TiO$_2$ substrate, in order to make a possible controllable magnetic 2D material on semiconductor substrate. We systematically investigate the structural, electronic, and magnetic properties of the heterostructure by first-principles calculations. Our calculated results show that the ground state of the heterostructure as a 2D electron system is an antiferromagnet-like semiconductor with a gap 0.4 eV, and when applying an electric field 0.026 V/\AA{}, it will transit to a half-metallic ferrimagnet with 100\% spin polarization. The more detailed results will be presented in the following.

\section{Computational details}

Our density-functional-theory calculations are done with a projector augmented wave (PAW) \cite{paw} method within the density functional theory \cite{dft1,dft2}, as implemented in the Vienna Ab initio Simulation Package (VASP) \cite{vasp1,vasp2}. We construct a uni-cell layer (one-unit-cell thickness) of TcO$_2$ on the rutile TiO$_2$ (001) surface. Our computational slab model consists of two Tc-O$_2$ monolayers, thirteen Ti-O$_2$ monolayers, and a vacuum layer with thickness of 20\AA. We take the lattice constants of the experimental bulk value of the rutile TiO$_2$ for the horizontal lattice constants of the slab, letting all the atoms relax for full optimization. We take the generalized-gradient approximation (GGA) of Perdew-Burke-Ernzerhof (PBE) version \cite{pbe} for the exchange-correlation functional. We consider electron correlation of Tc atoms by using GGA$+U$ method, taking $U = 2.3$ eV and $J = 0.3$ eV \cite{stc2}. For $\Gamma$-centered grids of $k$-points, $6\times 6\times 1$ is used to optimize the crystal structure of the slab model and $12\times 12\times 1$ is used to calculate the total energies of the slab model. The plane wave energy cutoff is set to 600 eV. Our convergence standard requires that the Hellmann-Feynmann force on each atomis less than 0.001 eV/\AA{} and the absolute total energy difference between two successive loops is smaller than $10^{-6}$ eV. The spin-orbit coupling is also taken into account to investigate the relativistic effects in the electronic structures. The direction of electric field is along $-z$ direction.

\section{Results and discussion}

\subsection{Field-driven phase transition and magnetic properties}

Rutile TiO$_2$ assumes a tetrahedral structure with lattice constants of 4.594 and 2.958 \AA{} \cite{tio2} and space group P4$_2$/mnm (\#136)\cite{co1,co2}. Bulk TcO$_2$ has the same symmetry as rutile TiO$_2$. In the TiO$_2$ (001) direction, the two adjacent Ti$^{4+}$O$_2$$^{2-}$ monolayers have a 90 degree rotation from each other. Our TcO$_2$/TiO$_2$ heterostructure is TcO$_2$  uni-cell epitaxial layer on rutile TiO$_2$ (001) surface. The slab model consists of TcO$_2$ uni-cell epitaxial layer (two Tc-O$_2$ monolayers), thirteen Ti-O$_2$ monolayers, and a vaccum of 20\AA{} thickness. The bottom surface is artificial, but fortunately, thanks to the structural optimization, there are no electronic states from the bottom TiO$_2$ surface in the energy window of -1.5 eV and 0.5 eV, which means that the thirteen Ti-O$_2$ monolayers in the slab model are enough to simulate the TiO$_2$ substrate in the heterostructure.

For zero electric field, the ground-state phase of the system is an antiferromagnet-like (AFM-like) semiconductor, having lower total energy than the ferrimagnetic (FiM) phase by 36 meV. The AFM-like phase is a semionductor, but it cannot be considered to be truly AFM because the surface Tc atom has a magnetic moment $1.96\mu_B$ and the subsurface Tc $-1.92\mu_B$, although the total magnetic moment is equivalent to zero. The metastable ferrimagnetic phase is half-metallic and has total moment $2\mu_B$, and the corresponding Tc moments are -0.37 and 1.94 $\mu_B$. Other magnetic phases are substantially higher in total energy.

\begin{figure}[!htbp]
{\centering  
\includegraphics[clip, width=8cm]{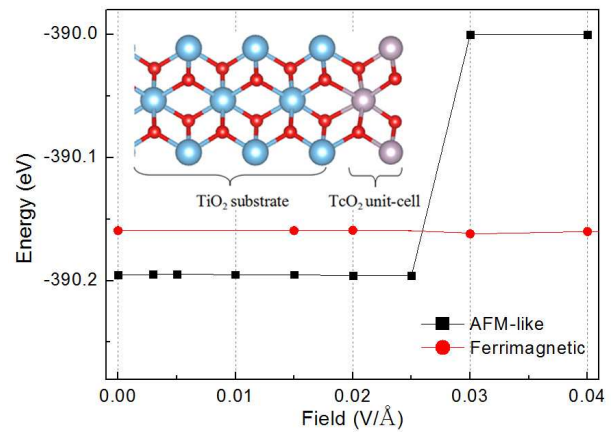}}\\
\caption{(Color online) The total energies of the AFM-like phase (black square) and the ferrimagnetic phase (red circle) of the TcO$_2$/TiO$_2$ heterostructure under different electric fields. Insert shows its structure consisting of TcO$_2$ uni-cell on rutile TiO$_2$ (001) surface, with the red, blue, and gray balls representing oxygen, titanium, and technetium atoms, respectively.}\label{fig1}
\end{figure}

After applying gate electric field, the magnetic phases the slab model are fully optimized, too. The total energy values of the AFM-like and ferrimagnetic phases under the electric fields up to 0.04 V/\AA{} are shown in Figure 1. Through the total energy comparison of the two phases, we find that the ground state of the system is still the AFM-like phase until the electric field is equivalent to 0.025 V/\AA, and then it transits to the ferrimagnetic phase when the electric field reaches 0.03 V/\AA. Therefore, the phase transition occurs when the applied electric field is between 0.025V/\AA{} and 0.03V/\AA. It is reasonable to define the transition point at 0.026 V/\AA{}.

It is interesting to investigate the effect of electric field on the electronic and magnetic properties of the two phases of the heterostructure. We summarize the semiconductor gap of the AFM-like phase and the half-metallic gap of the ferrimagnetic phase for different electric fields in Table I. For the AFM-like phase, it is a semiconductor up to the field 0.04 V/\AA{}, and the ferrimagnetic phase is still half-metallic up to 0.04 V/\AA{}. The stable phase, with the lowest total energy, is the AFM-like semiconductor of gap 0.4 eV up to 0.025 V/\AA{}, and then it transits to the ferrimagnetic half-metal at 0.03 V/\AA{}. In addition, we also present the total magnetic moments and partial moments in the Tc muffin tins of the two phases in Table I. With the electric fields, the total magnetic moment remains to be 0 for the AFM-like phase, and the ferrimagnetic phase always has 2$u_B$, indicating half-metallicity. The partial Tc magnetic moments change little or a little with the electric field.

\begin{table}[!h]
\caption{The semiconductor gap ($G_s$), total magnetic moment ($M_t$), and partial moments of the surface Tc ($m_1$) and subsurface Tc ($m_2$) of the AFM-like phase, and the half-metal gap ($G_h$) and the corresponding magnetic moments of the ferrimagnetic phase, of the TcO$_2$/TiO$_2$ heterostructure for different electric fields ($E$).}
\begin{ruledtabular}
\begin{tabular}{ccccccc}
$E$ (V/\AA) &  0  &  0.015 &  0.02  &  0.025  &  0.03   & 0.04  \\ \hline
$G_s$ (eV)  &  0.4    &  0.4   &  0.4   &  0.4    &  0.12   &  0.13  \\
$M_t$ ($\mu_B$) & 0   &  0     &  0     &   0     &  0      &  0    \\
$m_1$ ($\mu_B$) & 1.96 & 1.96  & 1.96   & 1.96  & -1.86   & -1.86  \\
$m_2$ ($\mu_B$) & -1.92 & -1.92  & -1.92   & -1.92  & 1.83   & 1.83  \\
\hline
$G_h$ (eV)  &  0.11   &  0.11  &  0.19  &   -     &  0.18   & 0.12 \\
$M_t$ ($\mu_B$) & 2   &  2     &  2     &   -     &  2      &  2    \\
$m_1$ ($\mu_B$) & -0.37 & -0.38  & -0.38 &  -  & -0.37   & -0.38  \\
$m_2$ ($\mu_B$) & 1.94 & 1.94  & 1.94   &   -  &  1.94   &  1.94  \\

\end{tabular}
\end{ruledtabular}
\end{table}

\subsection{Field-dependent electronic structures}

To show the effect of the phase transition on the electronic structures of the stable phases, we present in Figure 2 the spin-resolved band structures of the heterostructure in the presence of zero field and 0.03 V/\AA. It is clear that the heterostructure is a typical semiconductor with a gap 0.4 eV. After taking the spin-orbit interaction into account, the gap becomes a little smaller. With an electric field 0.03 V/\AA{} applied, the band structure of the heterostructure transits to a half-metallic ferrimagnet, changing little after taking SOC into account.

\begin{figure}[!htbp]
{\centering  
\includegraphics[clip, width=8.4cm]{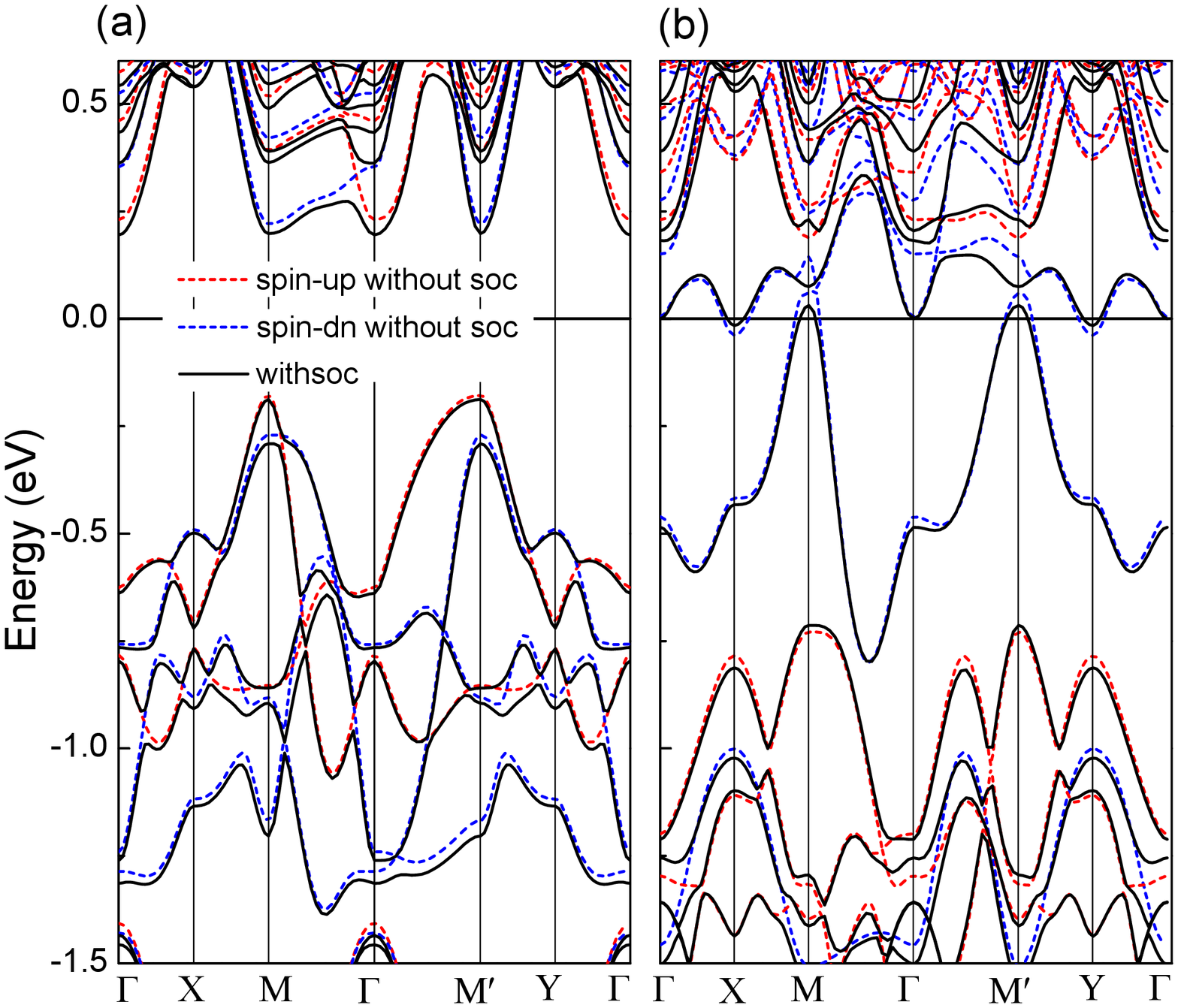}}\\
\caption{(Color online) Spin-resolved band structures of the TcO$_2$/TiO$_2$ heterostructure at the electric fields: $E=0$ (a) and $E=0.03$ V/\AA{} (b). The red dash lines and blue dash lines represent the bands of spin up and spin down without SOC, and the black solid lines show the band structure with SOC taken into account.}\label{fig2}
\end{figure}

\begin{figure}[!htbp]
{\centering  
\includegraphics[clip, width=8.4cm]{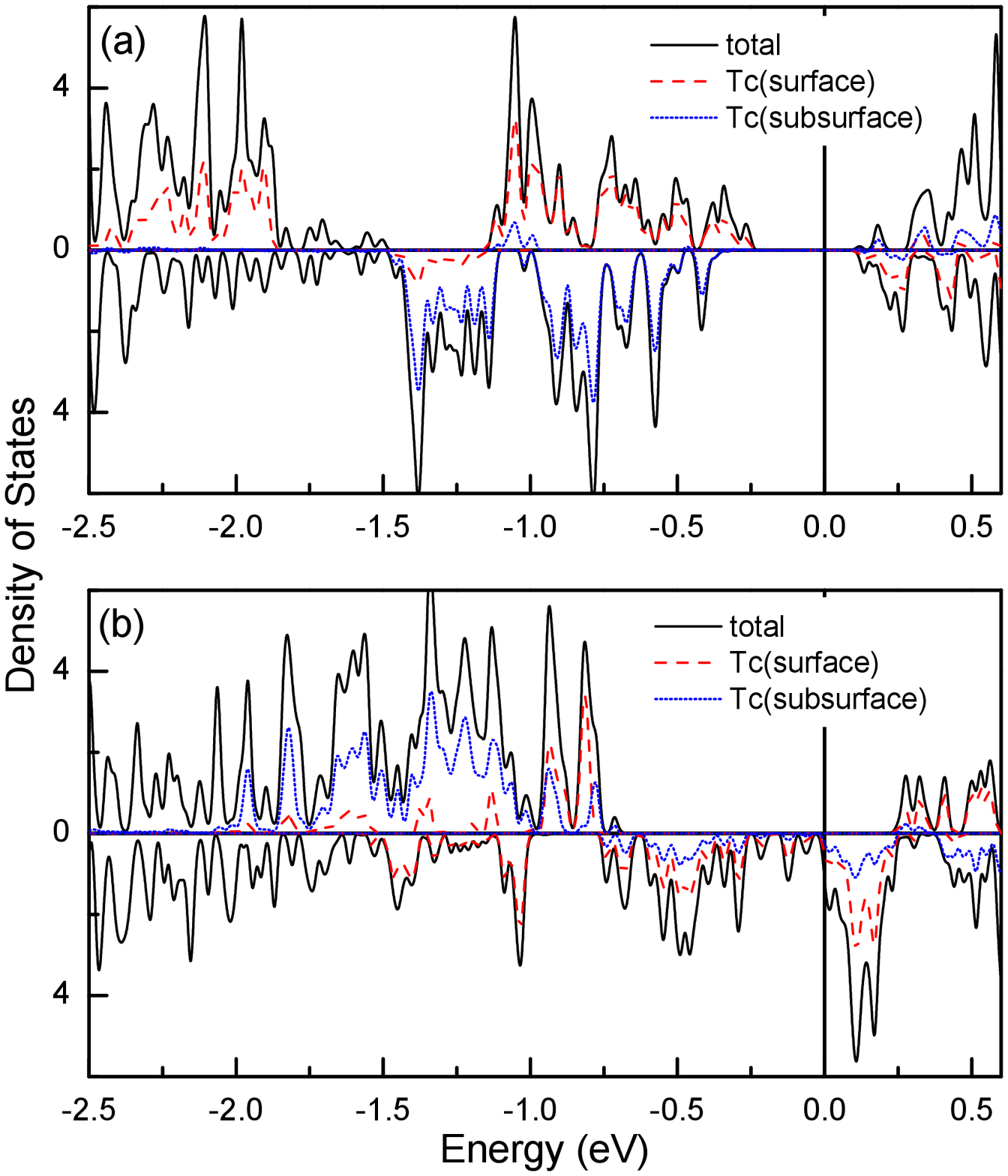}}\\
\caption{(Color online) Spin-resolved density of states (DOS) of the TcO$_2$/TiO$_2$ heterostructure with SOC at the electric fields $E=0$ (a) and $E=0.03$ V/\AA{} (b). The black lines, red dash lines, and blue dot lines represent the total DOS, the partial DOS of surface Tc-d, and the partial DOS of subsurface Tc-d, respectively. The upper and lower parts of each panel represent the DOSs of spin up and spin down, respectively.}\label{fig3}
\end{figure}

In Figure 3 we present the corresponding spin-resolved densities of states (DOSs) of the TcO$_2$/TiO$_2$ heterostrucutre, with SOC taken into account. For zero field, because the stable phase is the AFM-like semiconductor, the DOS of the two spin channels are different from each other, in contrast to symmetrical DOS of usual AFM semiconductor, and the semicoductor gap is made by the spin-up valence band top and the spin-down conduction band bottom. The DOSs of spin-up and spin-down channels between -1.5 eV and 0 are mainly from surface Tc atom and subsurface Tc atom, respectively. For 0.03 V/\AA, the DOS shows  typical half-metallic feature. There is a gap of 1 eV in the spin-up DOS, and the spin-down DOS is metallic.
As detailed analysis shows, it is independent of the applied electric field that the DOS in the energy window between -1 to 0 eV originate from the Tc-d orbital of the surface and subsurface monolayers. The DOS around the Fermi energy is due to the TcO$_2$ layer and the top two Ti-O$_2$ monolayers. At zero field, the surface and subsurface Tc-O$_2$ monolayers contribute almost equally to DOS, but for 0.03 V/\AA{} the subsurface Tc-O$_2$ monolayer plays an major role in the DOS of the occupied states. It can be seen that the main electronic states near the Fermi level are  from the TcO$_2$ layer and the remaining minor part originates from the two interfacial TiO$_2$ monolayers. Consequently, the heterostructure hosts a 2D electron system, which is very promising because it is a narrow-gap semiconductor when the electric field is less than 0.026 V/\AA{} and transits to a half-metallic ferrimagnet with 100\% spin-polarization at 0.026 V/\AA{}.

\subsection{Field-dependent structure parameters}

We present in Figure 4 the monolayer-resolved M-O (M=Tc, Ti) bond lengths ($l_h$, $l_s$) and O-O bond lengths ($l_O$) for zero field, 0.02 V/\AA{}, and 0.03 V/\AA{}. The surface monolayer is indicated by '1', the subsurface monolayer by '2', and so on. $l_h$  describes the horizontal M-O bond, $l_s$  corresponds to the skew M-O bond, and $l_O$ is the bond length of the horizontal O-O dimer. It is clear that for the surface and subsurface monolayers, the three bond lengthes substantially deviate from the corresponding values of the bulk TiO$_2$ phase (the horizontal dash lines). As an approximate rule, the stronger the electric field is, the larger the deviation becomes. The deviation tends to decrease when crossing the interface and entering the TiO$_2$ layer. When the number of monolayer  increases to 7, however, all the three bond lengthes tend to converge to the corresponding bulk values.

\begin{figure}[!htbp]
{\centering  
\includegraphics[clip, width=7.5cm]{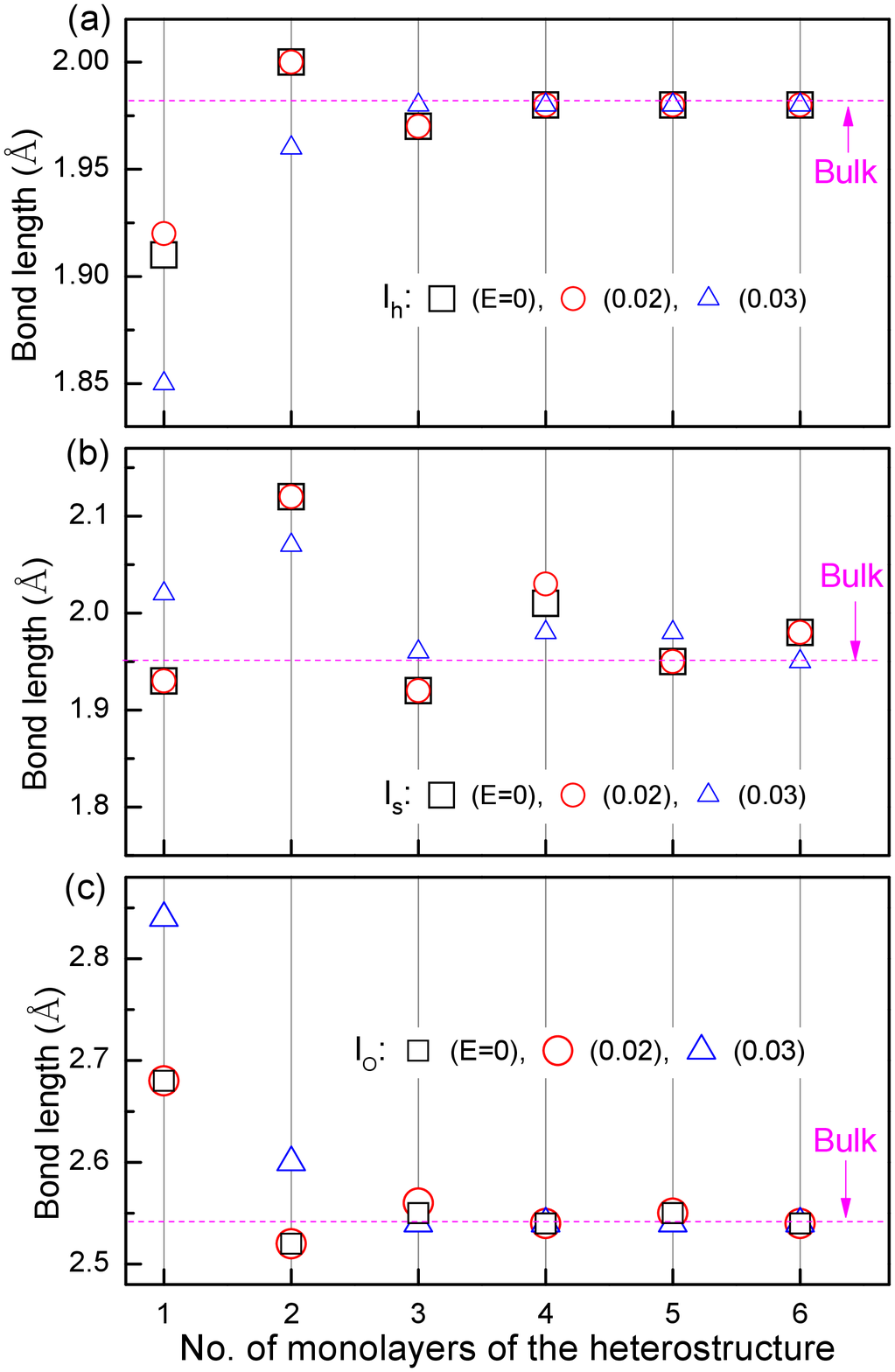}}\\
\caption{(Color online) The horizontal M-O bond lengths ($l_h$) (a), ckew M-O bond lengths ($l_s$) (b), and the horizontal O-O bond length ($l_O$) (c) of the TcO$_2$/TiO$_2$ heterostructure under different electric fields. The squares, circles, and triangles represent the data for 0, 0.02, and 0.03 V/\AA, respectively. 'Bulk' indicates the corresponding values of the bulk TiO$_2$ crystal.}\label{fig4}
\end{figure}

We present the monolayer-resolved bond angles ($\alpha$ and $\beta$) of the TcO$_2$/TiO$_2$ heterostructure in Table II. The subscripts '1', '2', and '3' indicate the number of the unit-cell layers from the top (surface). Both $\alpha$ and $\beta$  describe the O-M-O (M=Tc and Ti) bond angles, but the two O atoms in $\alpha$ belong to the same monolayer and those in $\beta$ are from two different monolayers. It is clear that $\alpha_i$ is smaller than 90$^\circ$, but $\beta$ is larger than 90$^\circ$. For the bulk rutile TiO$_2$ phase, $\alpha_B$ is 81.0$^\circ$ and $\beta_B$ is equivalent to 90$^\circ$.
It is very interesting that both $\alpha_1$ and $\beta_1$ of the top unit-cell layer substantially deviate from the corresponding bulk values, but $\alpha_i$ and $\beta_i$ converge to the bulk values when $i$ becomes larger than 3.

\begin{table}[!h]
\caption{The bond angles ($\alpha$ and $\beta$ in $^\circ$) of the TcO$_2$/TiO$_2$ heterostructure in the first (1), second (2), and third (3) unit-cell-layers under the fields: $E=0$, 0.02, and 0.03 V/\AA. The bulk values are indicated with the subscript 'B'.}
\begin{ruledtabular}
\begin{tabular}{ccccccccc}
$E$&  $\alpha_1$  & $\alpha_2$ &  $\alpha_3$ & $\alpha_B$ & $\beta_1$  &   $\beta_2$  &  $\beta_3$ & $\beta_B$\\ \hline
0     &  87.9  & 83.5  &  81.0  & 81.0 &  94.5  &  92.0  &  90.5  & 90  \\
0.02  &  87.9  & 83.6  &  81.0  & -    &  94.5  &  92.0  &  90.5  & -  \\
0.03  &  89.4  & 80.7  &  80.0  & -    &  94.8  &  91.2  &  89.8  & -
\end{tabular}
\end{ruledtabular}
\end{table}

\begin{figure}[!htbp]
{\centering  
\includegraphics[clip, width=8cm]{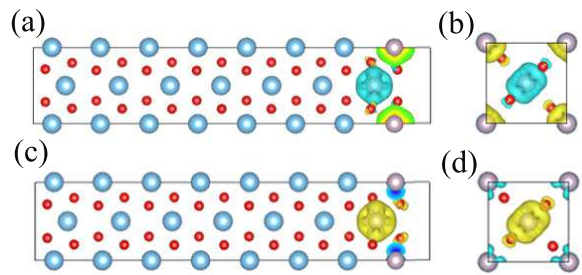}}\\
\caption{(Color online) The magnetization density of the heterostructure in the energy window of -2 to 0 eV under zero field (a,b) and 0.03 V/\AA{} (c,d). (a) and (c) represent the side views, and (c) and (d) represent the top views. The yellow and cyan isosurfaces ($\pm$0.01${|e|}$/\AA$^3$) represent positive and negative magnetization density, respectively.}\label{fig5}
\end{figure}

\subsection{Further discussions}

In order to show the real-space feature of the magnetic property in the TcO$_2$/TiO$_2$ heterostructure, we present in Figure 5 the magnetization density distribution (from the energy window between -2 and 0 eV) of the whole heterostructure for zero field and 0.03 V/\AA{}. It is clear that the finite magnetization density is located to the two Tc-O$_2$ monolayers, which is consistent with the 2D electron feature of the heterostructure. At zero filed, the surface Tc atom have positive magnetization density value, and the subsurface Tc atom has negative one. It can be seen that the absolute values of these two magnetization densities are almost equal to each other. When the electric field reaches to 0.03 V/\AA{}, it is clear that the magnetization density of the surface Tc atom and the subsurface Tc atom become substantially different. The subsurface Tc atom contributes almost the same value as that for zero field, but the surface Tc atom contributes much smaller than that for zero field. It clearly indicates that the heterostructure transits to the ferrimagnetic phase when the electric field changes from zero to 0.03 V/\AA{}, which is consistent with the partial magnetic moments in Table I.

As for the magnetocrystalline anisotropy energy of the stable magnetic phase, our calculated results show that it is approximately equivalent to 2 meV and the easy axis is in the plane for the AFM-like phase (zero field), but it is 67 meV and the easy axis is along the $z$ axis for the ferrimagnetic phase (electric field 0.03 V/\AA). It is a big challenge to directly estimate the Curie temperature of the ferrimagnetic phase (electric field 0.03 V/\AA). Considering that the easy axis is along the $z$ axis and both the CaTcO$_3$ and SrTcO$_3$ have very high transition temperatures (exceeding 800 K and 1000 K)\cite{stco1,ctco1}, we believe that the magnetic phase transition temperature of the ferrimagnetic phase of the TcO$_2$/TiO$_2$ heterostructure should be high enough to maintain the key magnetic properties at least up to room temperature.

\section{Conclusion}

Through first-principles optimization and calculation, we have investigated the structural, magnetic, electronic properties of the TcO$_2$ uni-cell layer on rutile TiO$_2$ (001) substrate as a TcO$_2$/TiO$_2$ heterostructure under various external electric field. It is shown that the heterostructure is a narrow-gap semiconductor with an AFM-like magnetic ordering when the applied electric field is less than 0.026 V/\AA{}, and at electric field 0.026V/\AA{} it transits to a half-metallic ferrimagnet with 100\% spin polarization. Our analysis indicates that  the field-driven phase transition is actually a double transition: from semiconductor to half metal, and from AFM-like order to ferrimagnet. The magnetization density and the electronic states near the Fermi level originate mainly from the Tc uni-cell layer, with the remaining minor part from the interfacial Ti-O$_2$ monolayers. It is reasonable to believe that the magnetic properties can be maintained at high temperature because Tc-based perovskite materials can have Curie temperatures around 1000 K. The bonds and bond angles quickly converge to the corresponding values of bulk TiO$_2$ when crossing the interface and entering the TiO$_2$ layer. Therefore, the heterostructure is actually a 2D electron system, determined by the Tc uni-cell layer and the TiO$_2$ substrate. Because the half-metallic phase has 100\% spin polarization and the transition field is easily achievable, this epitaxially obtainable 2D electron system could be used in spintronics applications.

\begin{acknowledgments}
This work is supported by the Nature Science Foundation of China (Grant No. 11574366), by the Strategic Priority Research Program of the Chinese Academy of Sciences (Grant No.XDB07000000), and by the Department of Science and Technology of China (Grant No. 2016YFA0300701). All the numerical calculations were performed in the Milky Way \#2 Supercomputer system at the National Supercomputer Center of Guangzhou, Guangzhou, China.
\end{acknowledgments}

\end{document}